\documentclass[pra, twocolumn, article,nofootinbib]{revtex4-1}

\usepackage{amsmath,amssymb,amsthm}
\usepackage{quantum}
\usepackage{color}
\usepackage{tikz}
\usepackage[justification=raggedright]{caption}
\usepackage{subcaption}
\usepackage{hyperref}

\begin{document}
\title{Tensor-Network Simulations of the Surface Code under Realistic Noise}

\author{Andrew S. Darmawan}
\affiliation{D\'epartement de Physique \& Institut Quantique, Universit\'e de Sherbrooke, Qu\'ebec, Canada}
\author{David Poulin}
\affiliation{D\'epartement de Physique \& Institut Quantique, Universit\'e de Sherbrooke, Qu\'ebec, Canada}

\begin{abstract}
    The surface code is a many-body quantum system, and simulating it in generic conditions  is computationally hard. While the surface code is believed to have a high threshold, the numerical simulations used to establish this threshold are based on simplified noise models. We present a  tensor-network algorithm for simulating error correction with the surface code under arbitrary local noise. We use this algorithm to study the threshold and the subthreshold behavior of the amplitude-damping and systematic rotation channels. We also compare these results to those obtained by making standard approximations to the noise models.
\end{abstract}

\date{\today}

\maketitle

\noindent{\em Introduction. --- }
The working principle behind quantum error correction is to ``fight entanglement with entanglement,'' i.e., protect the data against local interaction with the environment by encoding them into delocalized degrees of freedom of a many-body system. Thus, characterizing a fault-tolerant scheme is ultimately a problem of quantum many-body physics.

While simulating quantum many-body systems is generically hard, particular systems have additional structure that can be taken advantage of. For example, free-fermion Hamiltonians have algebraic properties that make them exactly solvable. An analogy in stabilizer quantum error correction is Pauli noise, where errors are Pauli operators drawn from some fixed distribution. Because of their algebraic structure, Pauli noise models can be simulated efficiently using the stabilizer formalism \cite{gottesman_stabilizer_1997}. Beyond Pauli noise, noise composed of Clifford gates and projections onto Pauli eigenstates can also be simulated efficiently using the same methods \cite{gutierrez_approximation_2013}. 

While such efficiently simulable noise models can be useful to benchmark fault-tolerant schemes, they do not represent most models of practical interest.  For instance, qubits that are built out of nondegenerate energy eigenstates are often subject to relaxation, a.k.a. amplitude damping. Miscalibrations often result in systematic errors corresponding to small unitary rotations  \cite{wallman_estimating_2015}. Given that these processes do not have efficient descriptions within the stabilizer formalism, understanding how a given fault-tolerant scheme will respond to them is a difficult and important problem.

The simplest approach to such  many-body problems is brute-force simulation, where an arbitrary state in Hilbert space is represented as an exponentially large vector of coefficients. 
Using such methods,  small surface codes (up to distance 3) have been simulated under non-Clifford noise \cite{tomita_low-distance_2014}. In another study, brute-force simulation of the seven-qubit Steane code  was performed without concatenation \cite{gutierrez_errors_2016-1}. Simulation of such low distance codes allows comparison of noise at the logical level to the noise on the physical level; however, it is difficult to infer  quantities of interest such as thresholds or overheads from such small simulations. Another approach, akin to the use of tight-binding approximations in solid-state physics, is to approximate these noise processes with efficiently simulable ones \cite{magesan_modeling_2013, gutierrez_approximation_2013}. However, the accuracy of these approximations can be very poor as we will show below.  

In this work, we import quantum many-body methods developed in the context of solid-state physics to study quantum error correction with realistic, non-Clifford noise models. Our construction hinges on the fact that the surface code is a projected-entangled-pair state (PEPS) with low bond dimension \cite{schuch_peps_2010}.  As a basic demonstration, we use our method to simulate the surface code under two non-Clifford local noise models: amplitude damping and systematic rotation. We assume that syndrome measurements are performed perfectly. We obtain thresholds for these noise models and also study error correction in the region of practical interest, where the noise strength is low relative to the threshold. These results are compared with those obtained using standard Pauli approximations, and significant discrepancies are observed. We have performed exact simulations on codes of up to 153 data qubits, while in contrast, previous studies using brute-force simulations were limited to 13 data qubits and 12 syndrome qubits ~\cite{tomita_low-distance_2014}. 

\medskip
\noindent{\em Tensor networks.} --- 
For our purposes, a tensor $A_{i_1, i_2, \dots, i_n}$ is an $n$-index array of complex numbers, where $i_k$ runs from $0$ to $D_k-1$, where $D_k$ is called the bond dimension of $i_k$. We represent a tensor graphically as a box and each index of the tensor as an edge emanating from that box. When two tensors are linked by an edge, the connected indices are summed over; for instance, 
\begin{center}
\includegraphics[scale=0.8]{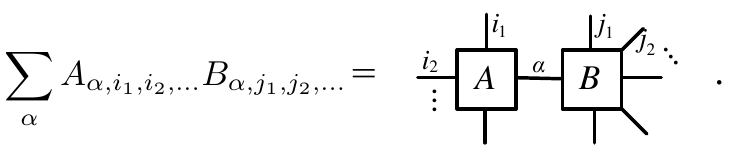}
\end{center}
It will often be convenient to group multiple indices together, using the notation $A_{i_1, i_2,\dots, i_n}=A_{\boldsymbol{i}}$ where a bold subscript is taken to mean that the index is composed of multiple indices. 

We define an $N$-particle PEPS $\ket{\psi}$ to be a quantum state whose tensor of coefficients $\psi_{i_1,i_2 \dots i_N}=\braket{i_1, \dots, i_N}{\psi}$ is the contraction of a network of $N$ tensors $\mathcal{T}=\{A^{(1)}_{i_1,\boldsymbol{\alpha}_1}, A^{(2)}_{i_2,\boldsymbol{\alpha}_2}, \dots, A^{(N)}_{i_N,\boldsymbol{\alpha}_N}\}$, each of which has one physical index labeled $i$ and some number of virtual indices labeled $\boldsymbol{\alpha}$.  We assume that the physical particles have dimension $d$, that each virtual index has bond dimension $D$, and that the number of virtual indices per tensor in $\mathcal{T}$ is less than a constant $n$. This implies that the tensor-network description of $\ket{\psi}$ is memory efficient: While the tensor $\psi_{i_1,i_2 \dots i_N}$ has $d^N$ entries, each tensor in $\mathcal{T}$ has at most $d D^n$ entries and thus the whole set $\mathcal{T}$ can be specified with only $Nd D^n$ complex numbers.  

\medskip
\noindent{\em Surface code tensor network. --- }
We consider the optimized layout of the surface code introduced in \cite{bombin_optimal_2007}, where qubits are placed on the vertices of a $W\times L$ rectangular lattice and $x$-check operators $A_f=\prod_{i\in f} X_i$ and $z$-check operators $B_f=\prod_{i\in f} Z_i$ are defined on alternating faces of the lattice in a checkerboard pattern. This layout is illustrated in Supplemental Material \ref{s:optimizations}.
The logical qubit state $\ket 0_L$  is defined to be the simultaneous $+1$ eigenspace of all check operators and of the logical operator $\overline Z$, where $\overline Z$ is a string of $Z$'s along the left boundary of the code. Likewise, the state $\ket 1_L$ is fixed by all check operators but is a $-1$ eigenstate of  $\overline Z$; it can be obtained as $\overline X \ket 0_L$, where $\overline X$ is a product of $X$ operators along the bottom boundary of the code. Given that the product state $\ket{0}^{\otimes N}$ has nonzero overlap with $\ket{0}_L$, and that $\ket{0}^{\otimes N}$ is a $+1$ eigenstate of every $B_f$ and $\overline{Z}$, we have 
\begin{equation}
    \ket{0}_L\propto \prod_{f}\fr{2}(I+A_f)\ket{0}^{\otimes N}\,,
    \label{e:logical_zero}
\end{equation}
where $\fr{2}(I+A_f)$ is the projector onto the $+1$ eigenspace of $A_f$, and the product is taken over all $x$ checks.
 
Let $W_{i,i',\boldsymbol{\alpha}}(C)$ be a tensor with two physical indices $i,i'$ and an arbitrary number of virtual indices $\boldsymbol{\alpha}=\alpha_1,\alpha_2,\dots$, which depends on a $2\times2$ matrix $C$. Each index of $W(C)$ has bond dimension two, and the only nonzero entries of $W(C)$ are 
\begin{align}
    W_{i,i',\boldsymbol{0}}(C)=\delta_{i,i'}\quad {\rm and}\quad
    W_{i,i',\boldsymbol{1}}(C)=C_{i,i'},
    \label{e:w_tensors}
\end{align}
where $\boldsymbol{0} (\boldsymbol{1})$ means that all virtual indices are set to 0 (1), and the symbol $\delta_{i,i'}$ denotes the Kronecker delta. For convenience, we define $Q^\pm=W(\pm X)$ and $R^\pm=W(\pm Z)$. 

Consider the projector $\fr{2}(I+A_f)$ acting on particles 1,2,3, and 4 ordered clockwise around a face. It can easily be verified that the tensor $\braopket{i_1,i_2,i_3,i_4}{I+A_f}{i_1',i_2',i_3',i_4'}$ can be expressed as a contraction of four tensors
\begin{equation}
    \sum_{\alpha_1,\alpha_2,\alpha_3}Q^{+}_{i_1,i_1',\alpha_1}Q^{+}_{i_2,i_2' \alpha_1,\alpha_2}Q^{+}_{i_3,i_3' \alpha_2,\alpha_3}Q^{+}_{i_4,i_4' \alpha_3}\,.
    \label{e:tensor_projector}
\end{equation}
We remark that the tensor description for the projection onto the ${-}1$ eigenspace of $A_f$ is identical to Eq. \eqref{e:tensor_projector} but with any one of the $Q^{+}$ tensors replaced with $Q^{-}$. Projections onto eigenspaces of $B_f$ operators can be defined analogously by replacing all $Q$'s with $R$'s. 
The tensor network corresponding to the product of projectors in Eq. \eqref{e:logical_zero}, is obtained by overlapping the tensor projector in Eq. \eqref{e:tensor_projector} over the whole lattice.

The state $\ket{0}_L$ is then obtained by applying this projector to the state $\ket{0}^{\otimes N}$, which, in the tensor-network picture,  corresponds to fixing the second index to zero, thus effectively removing it. A square-lattice tensor network $\mathcal{T}=\{A^{(1)}_{i_1,\boldsymbol{\alpha}_1}, A^{(2)}_{i_2,\boldsymbol{\alpha}_2}, \dots, A^{(N)}_{i_N,\boldsymbol{\alpha}_N}\}$ for the state $\ket{0}_L$ is thus formed from contractions of $Q^{+}$ tensors.

While this tensor network describes the logical $\ket{0}_L$ state, we need to be able to represent other states in order to fully characterize the transformation of the encoded information during error correction. Specifically, we want to compute the logical channel $\mathcal{E}_L$ that is applied to the logical qubit during a round of error correction. By the Choi-Jamiolkowski isomorphism, this channel can be inferred directly from the resulting output when a Bell state of the form $\ket{\Psi^+}=\ket{0}_L\ket{0}_a+\ket{1}_L\ket{1}_a$ is input, where the first qubit is encoded in a surface code and is subject to error correction, while the second qubit is unencoded and assumed to be noise free. We can obtain a tensor-network description for $\ket{\Psi^{+}}$ by a simple modification to the tensor network describing $\ket{0}_L$. Consider the tensor  
\begin{equation}
    \overline{\mathrm{CNOT}}:=\sum_{\alpha_1,\alpha_2,\dots,\alpha_L}Q^{+}_{i_1,i_1',\alpha_1}Q^{+}_{i_2,i_2' \alpha_1,\alpha_2}\dots Q^{+}_{i_L,i_L' \alpha_L, \alpha_a}\,,
\end{equation}
which can be thought of as an $L$ qubit operator (expressed in tensor form) with a single uncontracted virtual index $\alpha_a$, which we call the ancilla index. If $\alpha_a$ is set to $0$, then $\overline{\mathrm{CNOT}}$ is simply the identity on its physical indices, while if $\alpha_a=1$, $\overline{\mathrm{CNOT}}$ is an $L$-fold tensor product of $X$. Therefore if $\overline{\mathrm{CNOT}}$ is applied to the bottom row of the tensor network for $\ket{0}_L$, the resulting tensor-network state $\psi_{i_1,i_2,\dots,i_N,\alpha_a}=\braket{i_1,i_2,\dots,i_N,\alpha_a}{\psi}$ (which includes the ancilla index as a physical particle) will be the desired Bell state $\ket{0}_L\ket{0}_a+(\overline{X}\ket{0}_L)\ket{1}_a=\ket{\Psi^+}$. 

\begin{figure}
    \includegraphics[width=0.5\textwidth]{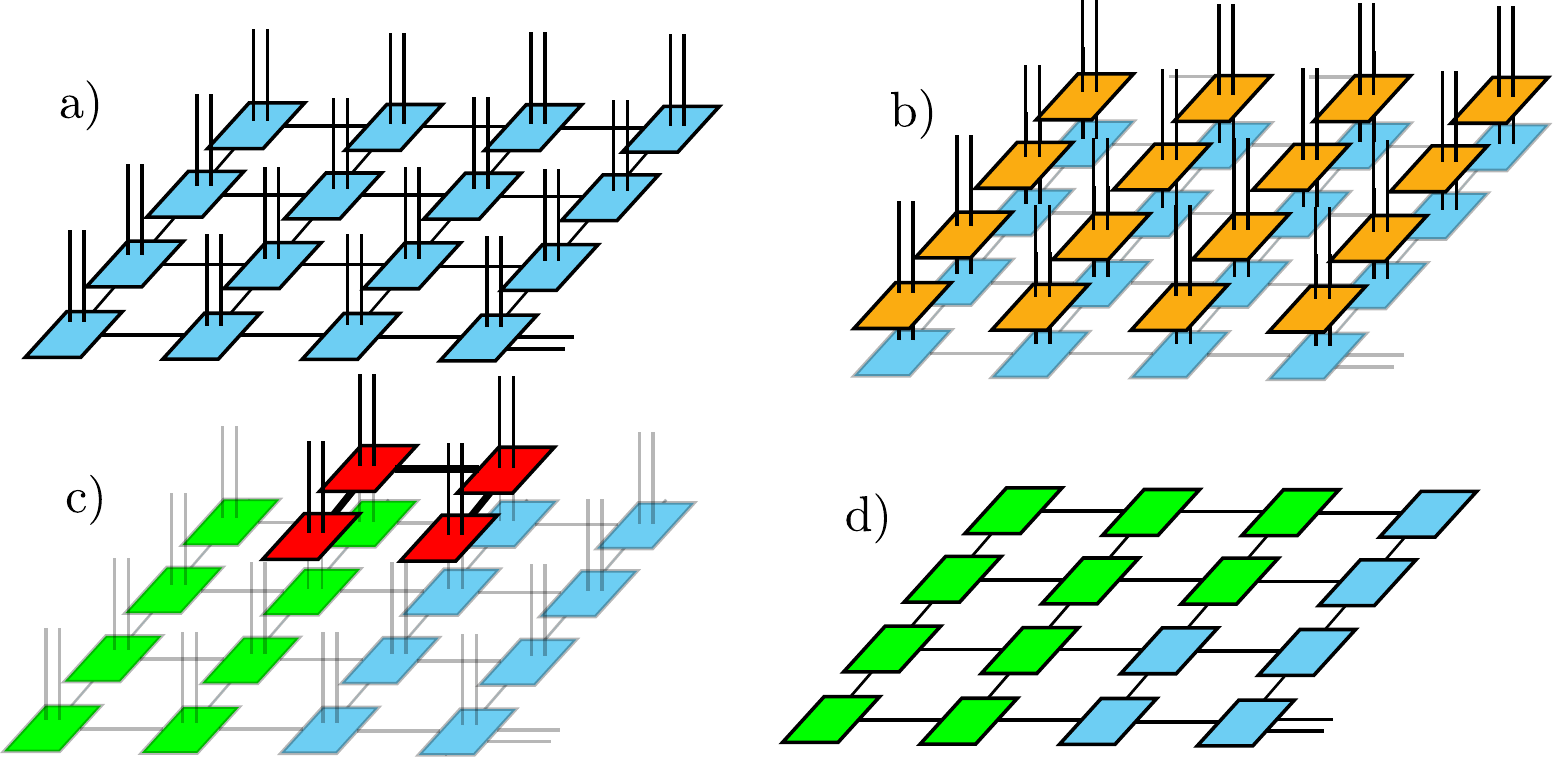}
    \caption{Simulation of error correction with the surface code. a) The surface-code density operator as a tensor network. The two additional indices on the bottom right tensor are ancilla indices. b) Applying local noise to the state. c) Applying a check projector $P_k$ (red) to the state. Green tensors have been involved in previous check measurements, while blue tensors have not. d) The resulting tensor network representing $Tr(P_{k+1}\rho_k)$ obtained by capping off pairs of physical indices. }
    \label{f:tn_evo}
\end{figure}

The above tensor-network definitions are for pure states. However, as we will be considering nonunitary noise, we want a tensor-network description for density matrices.  Given a tensor network $\{A^{(1)}_{i_1,\boldsymbol{\alpha}_1}, A^{(2)}_{i_2,\boldsymbol{\alpha}_2}, \dots, A^{(N)}_{i_N,\boldsymbol{\alpha}_N}\}$ for a pure state $\ket{\psi}$, we obtain the tensor network $\{B^{(1)}_{\boldsymbol{i_1},\boldsymbol{\alpha}_1}, B^{(2)}_{\boldsymbol{i_2},\boldsymbol{\alpha}_2}, \dots, B^{(N)}_{\boldsymbol{i_N},\boldsymbol{\alpha}_N}\}$ for the density matrix $\braket{i_1,\dots,i_N}{\psi}\braket{\psi}{i'_1,\dots,i'_N}$  by defining  $ B^{(k)}_{\boldsymbol{i},\boldsymbol{\alpha}}=A^{(k)}_{i,\boldsymbol{\alpha}'}A^{(k)*}_{i',\boldsymbol{\alpha}''}$,
where $\boldsymbol{i}=i,i'$ and $\boldsymbol{\alpha}=\boldsymbol{\alpha}',\boldsymbol{\alpha}''$ represent combined sets of physical and virtual indices respectively. The resulting tensor network is illustrated in Fig. \ref{f:tn_evo}a).

\medskip

\noindent{\em Simulation of error correction.} --- 
The evolution of the tensor network during the error-correction simulation is illustrated in Fig. \ref{f:tn_evo}. We start with the density matrix for the half-encoded Bell state described above. The noisy state is obtained by applying the desired CPTP map $\mathcal{E}$ to every qubit in the code. This corresponds to the local tensor update
\begin{equation}
    B_{i, i',\boldsymbol{\alpha}}\gets\sum_{j,j'}\mathcal{E}_{ijj'i'}B_{j, j',\boldsymbol{\alpha}}\,,
\end{equation}
where $\mathcal{E}_{ijj'i'}:=\bra{i}\mathcal{E}(\ketbra{j}{j'})\ket{i'}$, as depicted in Fig. \ref{f:tn_evo}(b).

During error correction, every check will be measured, yielding a set of measurement outcomes $s = (m_1,m_2,\dots)$ corresponding to a syndrome $s$ with probability $p_s=Tr(\Pi_s\rho)$,  where $\Pi_s$ is the projection onto the syndrome subspace corresponding to $s$ and $\rho$ is the noisy code state. We assume that measurements are performed perfectly. In order to sample from the distribution $p_s$ we imagine measuring checks sequentially. We can compute the conditional probability $p(m_k|m_1,m_2,\dots,m_{k-1})$ for any $k$ as follows. Let $\rho_k$ be the state of the system after $k$ check measurements and let $P_{k+1}$ be the projection onto the $+1$ eigenspace of the $k+1$ check. The probabilities of obtaining an outcome of $1$ and $-1$ are given, respectively, by $q=Tr(P_{k+1}\rho_k)$ and $1-q$. To obtain a tensor network for $P_{k+1}\rho_k$, we apply the tensor  corresponding to the check projector to the tensor network of $\rho_k$ as illustrated in Fig. \ref{f:tn_evo}c). For example, a four-qubit $x$ check can be decomposed as a contraction of four $Q^{+}$ tensors, as in Eq. \eqref{e:tensor_projector}. To apply the check to $\rho_k$ we contract these $Q^{+}$ tensors with the measured physical indices, specifically 
\begin{equation}
    B_{i, i',\boldsymbol{\alpha}} \gets \sum_{j} Q^+_{i,j,\boldsymbol{\alpha}'}B_{j,i',\boldsymbol{\alpha}''}\,,
    \label{e:apply_check_tensor}
\end{equation}
where the updated virtual indices $\boldsymbol{\alpha}$ contain the original virtual indices of $B$ as well as those of the appended $Q^+$.  Then, to evaluate $q$, we take the trace of every tensor ($B_{\boldsymbol{\alpha}}\gets \sum_i B_{i, i,\boldsymbol{\alpha}}$). The resulting tensor network has no physical indices and is illustrated in Fig. \ref{f:tn_evo}d). We then contract all virtual indices in the network. 

This last step corresponds to contracting a square-lattice tensor network. Generically, this problem is $\#P$ complete, and therefore, no efficient algorithm is believed to exist \cite{schuch_computational_2007}. However, many efficient algorithms have been developed to obtain approximate solutions to this problem \cite{verstraete_renormalization_2004, levin_tensor_2007, gu_tensor-entanglement-filtering_2009, lubasch_unifying_2014, evenbly_tensor_2015-1, yang_loop_2017}. In this work, we have used both an exact, albeit inefficient, contraction algorithm and an efficient, approximate contraction algorithm for finite-sized PEPS with open boundary conditions \cite{verstraete_renormalization_2004}.

To contract the tensor network exactly, we merge all tensors of the left column into a single tensor and then contract this sequentially with the remaining tensors from left to right. The algorithm is inefficient because the amount of memory required to store the tensor associated with a column is exponential in the lattice width. We note, however, that we have used a number of optimizations to significantly improve the efficiently of the algorithm. These are described in Supplemental Material \ref{s:optimizations}.  With exact contraction, the complexity of sampling a single syndrome is $\mathcal{O}(W4^W)$, and thus, the complexity of sampling all syndromes (and of the entire algorithm) is $\mathcal{O}(LW^24^W)$. For efficient, approximate contraction, we use a well-known algorithm described in \cite{schollwock_density-matrix_2011}, which represents the contraction of the network as the repeated multiplication of a matrix-product state by matrix-product operators. This algorithm depends on an accuracy parameter $\chi$ and contraction of the lattice takes time $\mathcal{O}(LW\chi^3)$.

When all check outcomes have been obtained, we have sampled a syndrome $s$ from the distribution $p_s$, and the tensor network encodes a noisy state with appropriate check projectors applied. The state is then returned to the code space via a Pauli operator, which we choose to be a product of $X$ operators connecting each flipped $B_f$ check to the left boundary and $Z$ connecting each flipped $A_f$ check to the top boundary.  

At this point, a classical algorithm, called the decoder, is used to select one of four possible corrections $\overline{I},\overline{X},\overline{Y},\overline{Z}$ to be applied to the state. The decoder uses information about the syndrome and the noise model to select the correction that minimizes the overall logical error. For efficiency, this is usually only done approximately \cite{dennis_topological_2002, harrington_analysis_2004, duclos-cianci_fast_2010, wootton_high_2012, bravyi_quantum_2013, duclos-cianci_fault-tolerant_2014, hutter_efficient_2014, watson_fast_2015,herold_cellular-automaton_2015, herold_cellular_2017}. In this work, we perform decoding by choosing the correction that minimizes the distance between the computed logical channel and the identity. The entire procedure (application of noise, check measurements, returning to the code space, correction) yields a CPTP  map $\mathcal{E}_L$ acting on the encoded qubit. The calculation of $\mathcal{E}_L$ as well as the decoding algorithm are performed using essentially the same method as the syndrome sampling and are described in detail in Supplementary Material \ref{s:logical_channel}.

\medskip
\noindent{\em Numerical results.} ---
Two non-Pauli noise models are considered: systematic rotation about the $z$ axis $\mathcal{E}_{SR}(\rho)=e^{-i\theta Z}\rho e^{i\theta Z}$, where $\theta\in[0,\pi)$ is the rotation angle, and amplitude damping $\mathcal{E}_{AD}(\rho)=\sum_i K_i \rho K_i^\dag$, which has two Kraus operators,
\begin{equation}
    K_0 = \ketbra{0}{0} + \sqrt{1-\gamma}\ketbra{1}{1}\,,\quad K_1 = \sqrt{\gamma}\ketbra{0}{1}\,,
\end{equation}
where $\gamma\in [0,1]$ is the damping parameter. As a test, we also performed simulations with the depolarizing channel $\mathcal{E}_{DP}(\rho)=(1-\epsilon)\rho + \frac{\epsilon}{3} X\rho X+ \frac{\epsilon}{3} Y\rho Y + \frac{\epsilon}{3} Z\rho Z$, which is a well-studied Pauli channel.

We have also considered two different Pauli approximations of these channels. The Pauli twirl approximation (PTA) to an arbitrary channel expressed in the Pauli basis $\mathcal{E}(\rho)=\sum_{i,j}\chi_{i,j}P_i \rho P_j$ is the channel $\mathcal{E}^{\rm PTA}=\sum_{i}\chi_{i,i}P_i \rho P_i$,  obtained by removing the off-diagonal elements of $\chi_{i,j}$ \cite{emerson_symmetrized_2007, dankert_exact_2009}. It can produce noise models that are much better behaved and thus provide poor insight into the performance of the real channel. For this reason, an honest Pauli approximation (HPA) was introduced \cite{magesan_modeling_2013}, which seeks the channel $\mathcal{E}^{\rm HPA}$, which is as close as possible to the original channel yet produces a noisier output on every possible input. This approximation is thus expected to provide a pessimistic bound on the performance of a fault-tolerant scheme under some noise process $\mathcal{E}$.

As a first application, we have used the exact simulation algorithm to estimate thresholds, which are presented in Table \ref{t:thresholds}. Details on how thresholds were determined are provided in Supplemental Material \ref{s:thresholds}. The largest lattice sizes simulated were $9\times 9 $ for depolarizing, $11\times 11$ for systematic rotation and $9\times 17$ for amplitude damping. For amplitude damping we found that a nonsquare lattice, where the $\overline{X}$ logical operator runs along the long dimension of the code, performed significantly better than a square one. 
\begin{table}
\begin{tabular}{|c|c|c|c|}
    \hline
    &Exact & PTA & HPA \\
    \hline
    DP ($\epsilon$) & $18.5\pm 1.5\%$ & $18.5\pm 1.5\%$ & $18.5\pm 1.5\%$ \\
    \hline
    AD ($\gamma$) &$39\pm2\%$ &$39\pm2\%$& $21\pm1\%$\\
    \hline
    SR ($\theta/\pi$)&$>0.15$&$0.17$&$0.055$\\
    \hline
\end{tabular}
    \caption{Computed thresholds for three noise models and their Pauli approximations. PTA and HPA to SR were computed using the exact value  of the threshold under bit flip noise. Pauli approximations of DP are identical to the exact channel, so only the exact channel was simulated.}
    \label{t:thresholds}
\end{table}

For the depolarizing channel, we can compare our obtained threshold with exact results. We find that our threshold estimate agrees with the optimal depolarizing threshold of $18.9(3)\%$\cite{bombin_strong_2012}. 

The systematic rotation channel was simulated for rotation angles in the range $0.025\pi\le \theta \le 0.3\pi$. A large region showed below threshold behavior; however, no clear transition  behavior could be identified. 

The thresholds of Pauli approximations did not always agree with the thresholds of the exact channels. However, the twirl approximation to amplitude damping, yielded the same threshold as the exact channel (to within the accuracy of our data). As expected, the honest Pauli approximations provided pessimistic values of the threshold for non-Pauli channels.

\begin{figure}[t]
         \includegraphics[width=9cm]{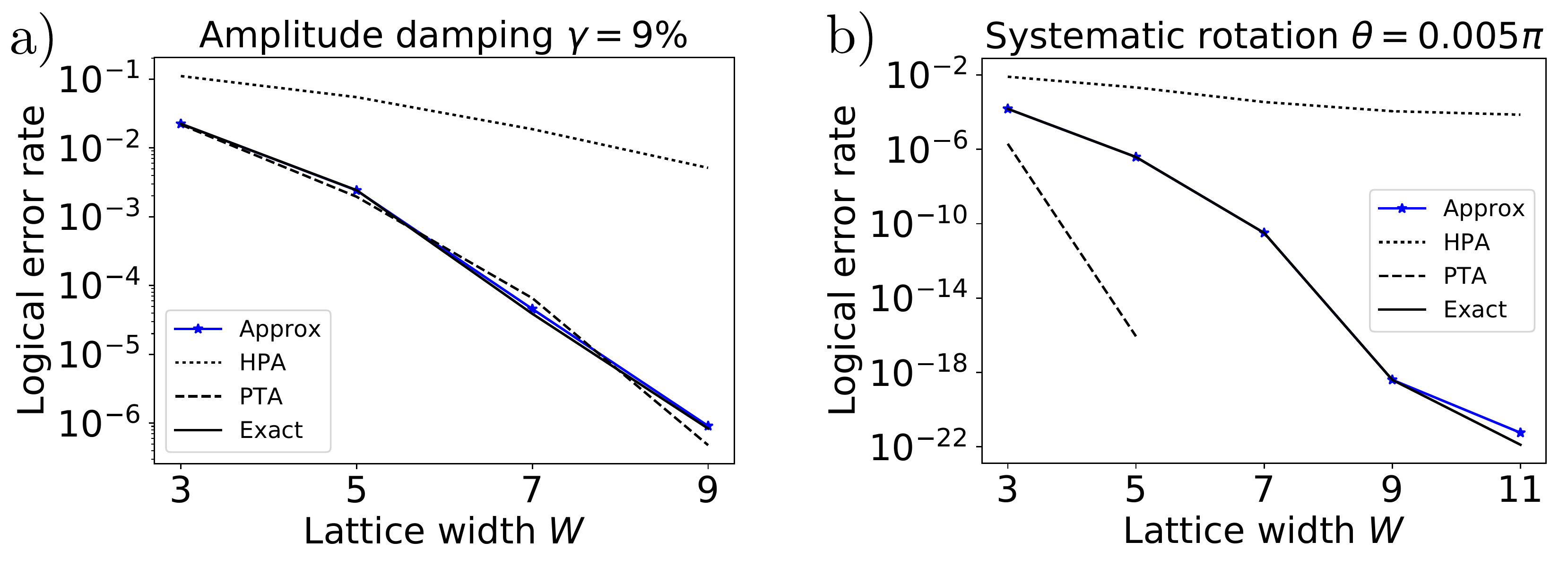}
     \caption{Empirical logical error rate vs lattice dimension for two non-Pauli noise models. Solid black lines represent exact channels while dashed and dotted lines represent the PTA and the HPA, respectively. The left plot is $\gamma=9\%$ amplitude damping, while the right plot is $\theta=0.005\pi$ systematic rotation. For each set of parameters, $1.32\times 10^5$ syndromes were sampled to acquire statistics. For the PTA of SR with $W>5$, error rates are numerically zero. Error rates obtained for the exact channels using an approximate contraction algorithm with $\chi=8$ are also plotted in blue. Note that the exact average logical error could significantly differ from the empirical average observed over millions of syndrome measurements if failures are dominated by outliers.}
     \label{f:below_threshold}
 \end{figure}

As another application of our algorithm, we have  performed exact simulations at fixed noise rates well below threshold. The results are presented in Fig. \ref{f:below_threshold} for amplitude damping with $\gamma=9\%$ and $z$ rotation with $\theta=0.005\pi$. Again, the behavior of the twirl approximation agrees quite well with that of the exact amplitude damping channel. However the behavior of both approximations differed considerably from the exact $z$ rotation. For instance, for a code with $W=5$, the PTA underestimates the logical error rate by a factor of about $10^{10}$, while the HPA overestimates the logical error rate by a factor of about $10^4$. The observed discrepancies highlight the fact that the performance of the code under an efficiently simulable approximation to a noise model can differ significantly from its performance under the exact noise model and motivates the development of more efficient algorithms, such as the one presented here.

We have also computed error rates using the exact channels but with an approximate contraction algorithm in place of the (inefficient) exact algorithm. Setting $\chi=8$ in the calculation of the logical channel, we observe remarkably good agreement with exact data. For instance, at high noise rates, we obtain the same threshold (within statistical error) for the amplitude damping channel (see the threshold plot in Supplemental Material \ref{s:thresholds}). We also observe very good agreement with exact data for systematic rotation and amplitude damping at low logical error rates, as is displayed in Fig. \ref{f:below_threshold}.

\medskip

\noindent{\em Conclusion.} ---
We have presented a simple tensor-network-based algorithm for simulating the surface code under arbitrary local noise. The algorithm can be made exact within statistical fluctuations, allowing accurate simulation of systems with well over 100 qubits. We have used exact simulation to estimate thresholds of non-Pauli noise models and to calculate error rates well below threshold.

In order to demonstrate scalability of the simulation, we have also used an approximate, efficient algorithm to calculate the logical channel in place of the exact algorithm. We found very good agreement with the exact data for non-Pauli noise at both high and low logical error rates.

While our simulation algorithm has assumed perfect check measurements, in practice, measurement errors are unavoidable. In surface-code error correction, measurement errors are detected by performing multiple rounds of check measurements and observing how the syndrome evolves in time. This procedure could potentially be simulated using PEPS time-evolution algorithms \cite{jiang_accurate_2008, pizorn_time_2011, orus_exploring_2012, werner_positive_2016} or by contracting 3D tensor networks \cite{xie_coarse-graining_2012}.

\medskip

\noindent{\em Acknowledgements. ---} 
We thank Marcus da Silva for discussions that initiated this project. This work was supported by the Army Research Office Contract No. W911NF-14-C-0048.

\bibliography{physics}
\bibliographystyle{apsrev4-1}

\pagebreak
\widetext
\begin{center}
    \textbf{\large Supplemental Material}
\end{center}
\setcounter{equation}{0}
\setcounter{figure}{0}
\setcounter{table}{0}
\setcounter{page}{1}
\makeatletter
\renewcommand{\theequation}{S\arabic{equation}}
\renewcommand{\thefigure}{S\arabic{figure}}
\renewcommand{\bibnumfmt}[1]{[#1]}
\renewcommand{\citenumfont}[1]{S#1}

\section{Computing the logical channel and decoding}
\label{s:logical_channel}
Here we provide more detail on how to compute the logical channel for a round of error correction. Note that, via the Choi-Jamiolkowski isomorphism, a qubit channel $\mathcal{E}$ is completely described by the $4\times4$ process matrix 
\begin{equation}
    C_{ij}=\mbox{Tr}([P_i\otimes P_j][(\mathcal{E}\otimes I) (\ketbra{\Psi^{+}}{\Psi^{+}})])\,,
\end{equation}
where $P_i$ for $i=0,1,2,3$ are the identity and Pauli X, Y and Z matrices respectively. This represents the state obtained when the channel is applied to the first qubit of a Bell state $\ket{\Psi^{+}}=\ket{00}+\ket{11}$, when expressed in the Pauli basis. Here we will show how to compute $C_{ij}$ for the logical channel in the case of surface-code error correction, which will depend on the noise map, the syndrome and the decoder. 

The simulated error correction process can be decomposed into three parts
\begin{equation}
    \mathcal{E}=\mathcal{D}_s\circ \mathcal{R}_s \circ \mathcal{N}\,,
\end{equation}
where $\mathcal{N}$ is the physical noise map acting on the $N$ qubits of the code, $\mathcal{R}_s$ is the recovery map and $\mathcal{D}_s$ is the decoder correction. The recovery map returns the noisy state to the code space (without performing any classical processing of the syndrome) and can further be decomposed into $\mathcal{R}_s(\rho)=T_s\Pi_s\rho \Pi_s T_s$, where $\Pi_s$ is a projector onto the subspace corresponding to the syndrome $s$ and $T_s$ is a Pauli operator which returns a state in the image of $\Pi_s$ to the codespace. We have defined $T_s$ as a product of Paulis that connects the flipped $X$ checks to the top boundary and flipped $Z$ checks to the left boundary. The decoder correction $\mathcal{D}_s$ is the correction selected from $\{\overline{I}, \overline{X}, \overline{Y}, \overline{Z}\}$ by the decoding algorithm to minimise the logical error. 

For now, assume that the decoder does nothing so that $\mathcal{E}=\mathcal{R}_s \circ \mathcal{N}$. The logical channel $\mathcal{E}$ is a map from the codespace to itself, so it is effectively a single-qubit map.
The process matrix describing the transformation of the logical information given the syndrome $s$ is thus
\begin{equation}
    C_{ij}=\mbox{Tr}([P_i\otimes P_j][((\mathcal{R}_s \circ \mathcal{N})\otimes I) (\ketbra{\Psi^+}{\Psi^+})])\,,
\end{equation}
where $\ket{\Psi^+}$ is a Bell state of the form $\ket{\Psi^+}=\ket{0}_L\ket{0}_a+\ket{1}_L\ket{1}_a$, where the first qubit is encoded into the surface code and the second qubit is an unencoded ancilla qubit that is assumed to be noise free.
Using the cyclic property of the trace and the fact that $T_s \Pi_s = \Pi_0 T_s$, where $\Pi_0$ is the projection onto the code space, we can write the above expression more explicitly as 
\begin{equation}
    C_{ij}=\mbox{Tr}([T_s P_i T_s \Pi_s\otimes P_j][ (\mathcal{N}\otimes I)(\ketbra{\Psi^+}{\Psi^+})])\,.
    \label{e:process_matrix}
\end{equation}
In the main text we have shown how to represent the Bell state as a tensor network, and how this tensor-network description changes as local noise $\mathcal{N}$, and syndrome projectors $\Pi_s$ are applied to the state. Using this, the trace over the physical indices only
\begin{equation}
    A_i=\mbox{Tr}_L([T_s P_i T_s \Pi_s\otimes I][ (\mathcal{N}\otimes I)(\ketbra{\Psi^+}{\Psi^+})])
    \label{e:partial_contraction}
\end{equation}
can be represented as a square lattice tensor network, as in \ref{f:tn_evo}d), and can be contracted using the method described in the text. Here $\mbox{Tr}_L$ indicates that the trace is taken over the physical indices of the logical qubit of $\ket{\Psi^+}$ and the ancilla qubit is left uncontracted, and untouched by the other operators in the trace. Therefore the expression in Eq. \eqref{e:partial_contraction} has two uncontracted indices (the ancilla indices), i.e. it is a $2\times2$ matrix. We use these free indices to compute the target quantities $C_{ij}$ in \eqref{e:process_matrix} via $C_{ij}=\mbox{Tr}(P_j A_i)$. 

Computing $C_{ij}$ in this way would seem to require computing four full square lattice tensor contractions: one for each $A_i$, i.e. each row of $C_{ij}$. However, as explained in Sec. \ref{s:optimizations} a lot of the intermediate contractions can be reused for different $i$, allowing the full matrix $C_{ij}$ to be computed using only two full lattice contractions. The overall complexity of this calculation is $\mathcal{O}(LW4^W)$ if performed exactly, or $\mathcal{O}(LW\chi^3)$ using the approximate algorithm.
Having computed the effective logical channel $\mathcal{R}_s\circ\mathcal{N}$, a decoder correction $\mathcal{D}_s$ can be incorporated simply by composing the logical channel with the chosen Pauli operator. 

Here we have shown how to compute the logical channel for a given syndrome and noise model. In the following section we will describe a decoding algorithm based on the above calculations.

\subsection{Exact decoding}

In the previous section we showed how to compute the logical channel $\mathcal{E}$ for a given noise model and syndrome. This same calculation allows us to perform exact decoding i.e. to select a optimal correction for a given syndrome. Assume that we are given a syndrome $s$ and a noise model $\mathcal{N}$. We can compute $\mathcal{R}_s \circ \mathcal{N}$ using the method described above. Then computing the optimal correction simply amounts to choosing the $\mathcal{D}_s$ from $\{I, X, Y, Z\}$ that minimises the distance of the effective channel from the identity 
\begin{equation}
    d({D}_s\circ \mathcal{R}_s \circ \mathcal{N},I)\,,
\end{equation}
where $d$ can be any distance between operators e.g. diamond distance, fidelity, or 2-norm distance. In this work we used the 2-norm distance due to its easy computability. If $\mathcal{R}_s \circ \mathcal{N}$ is calculated exactly, this decoding algorithm is optimal, in the sense that it exactly chooses the correction that minimises the distance of the logical channel from the identity. It involves the exact calculation of $\mathcal{R}_s \circ \mathcal{N}$ and is therefore exponential, as described above. However, if $\mathcal{R}_s \circ \mathcal{N}$ is already calculated, the remaining calculations involve manipulations of small matrices and take constant time. Thus choosing an optimal correction requires negligible extra work compared to exactly simulating error correction, and therefore we have used this above optimal decoding in all of our exact simulations.  When the approximate contraction algorithm is used, the decoder functions almost identically, except that it uses an approximation of $\mathcal{R}_s \circ \mathcal{N}$, rather than the exact channel. In this case, due to possible errors in the approximation, the decoding is not guaranteed to be optimal. However it would be interesting to see whether such a decoding algorithm is sufficiently accurate for practical purposes.

\section{Computing the error threshold}
\label{s:thresholds}
\begin{figure}[h]
    \centering
    \begin{subfigure}[b]{0.3\textwidth}
        \includegraphics[width=\textwidth]{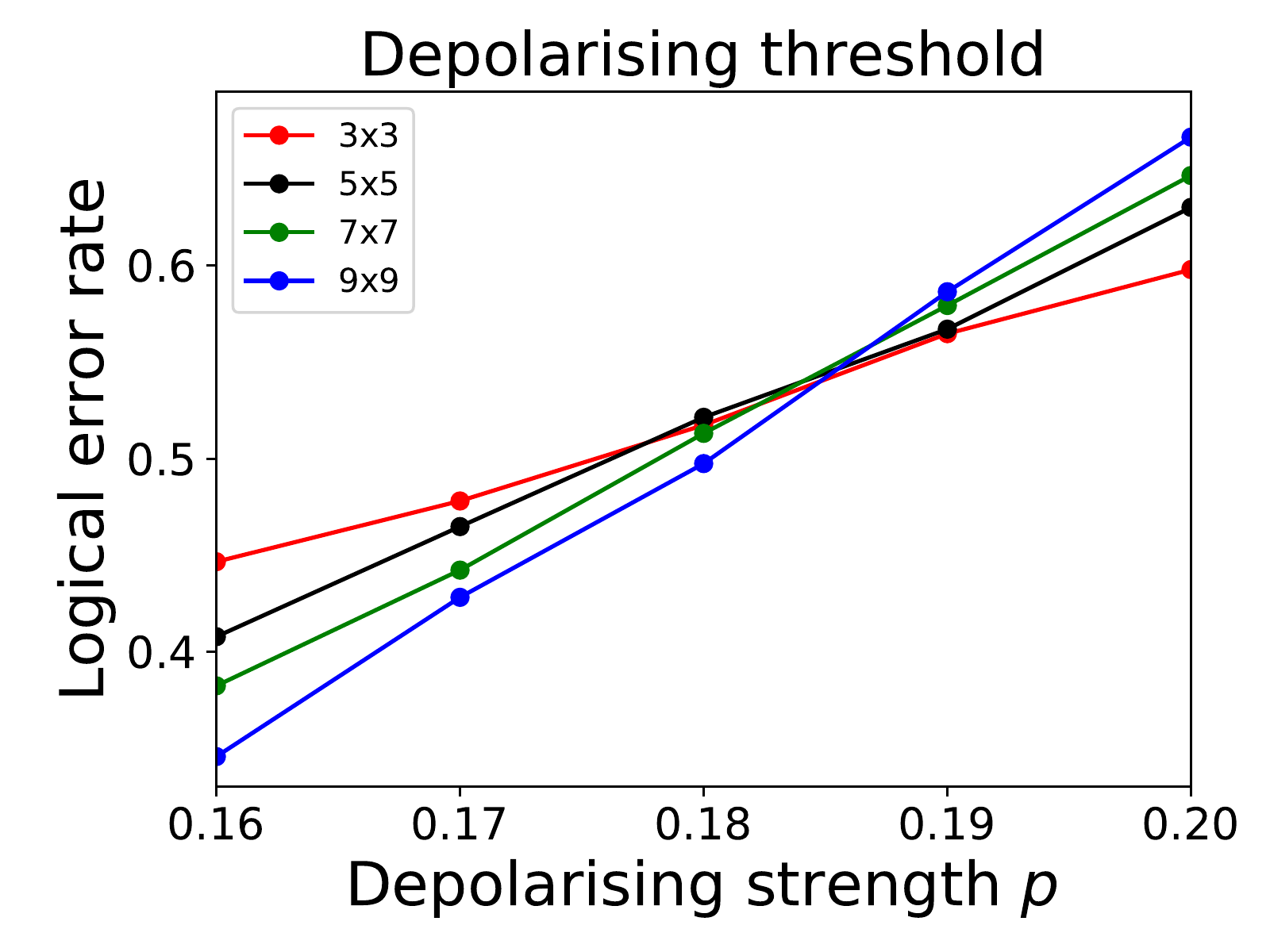}
    \end{subfigure}%
    \begin{subfigure}[b]{0.3\textwidth}
        \includegraphics[width=\textwidth]{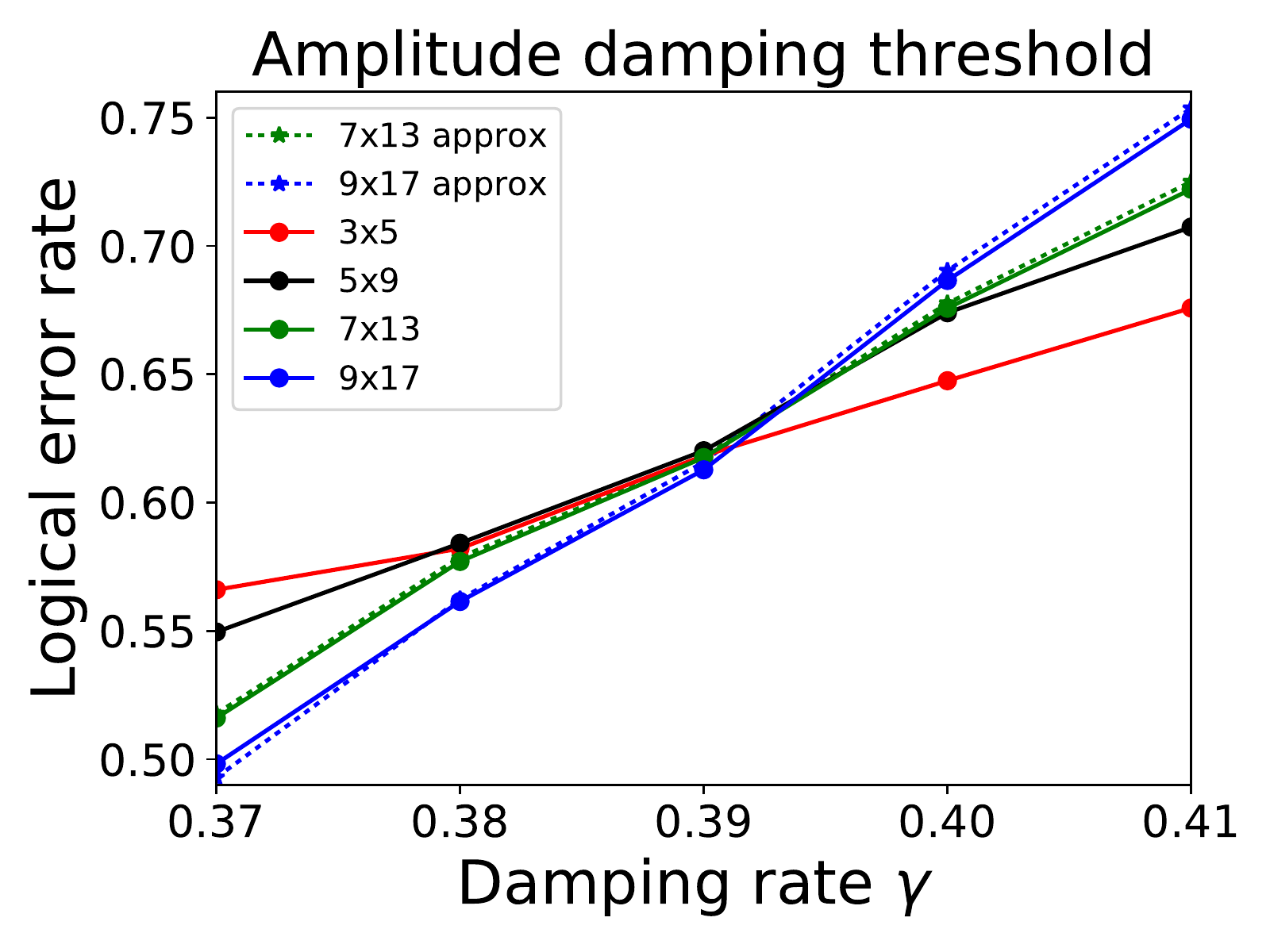}
    \end{subfigure}%
    \begin{subfigure}[b]{0.3\textwidth}
        \includegraphics[width=\textwidth]{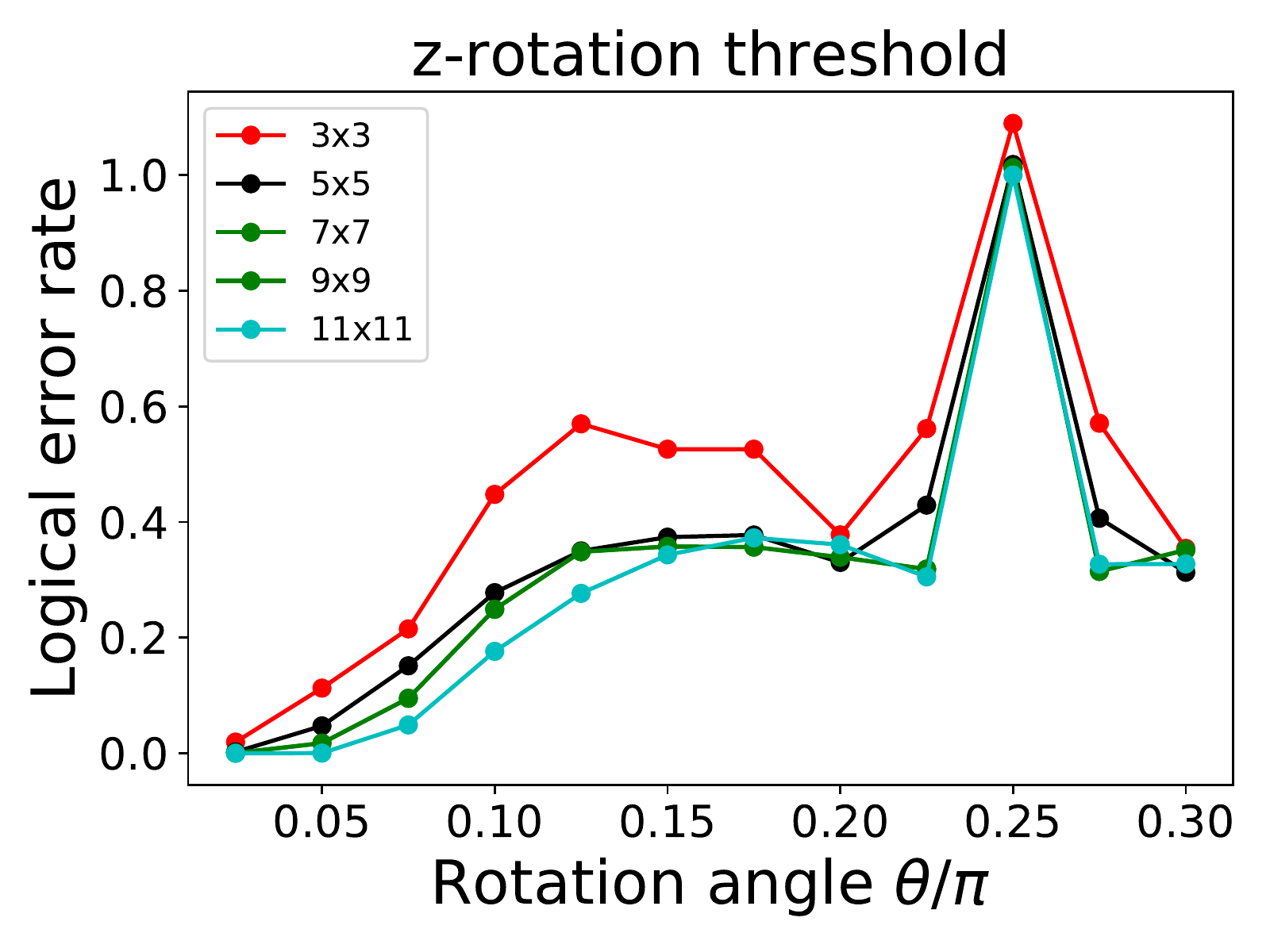}
    \end{subfigure}
    \caption{Logical error rate (defined as the diamond distance of the logical channel from the identity) vs. noise strength for various noise models and lattice sizes.}
    \label{f:thresholds}
\end{figure}

In Fig. \ref{f:thresholds} we include plots obtained from our simulation data. The simulations were run for depolarizing, amplitude damping, and systematic rotation noise models and the lattice dimensions and the noise strength were varied. For each data point, a large number of samples (between $1.3\times 10^4$ and $1.3 \times 10^5$) were obtained. We have used the decoding algorithm described in the previous section. A threshold was identified as the point below which increasing the lattice size resulted in a decrease in the logical error rate. For amplitude damping we have also plotted data obtained using an approximate contraction algorithm to compute the logical channel for the largest two lattice sizes. 

For depolarizing and amplitude damping, the thresholds are clear from the graphs and are listed in Table \ref{t:thresholds}. We remark that, for amplitude damping, the crossing point for the approximate data is identical to that obtained using exact data. On the other hand, the z-rotation did not have a clear threshold. The plot is symmetric about the point $\theta=0.25\pi$, where the code performs most poorly. Increasing the lattice size appeared to decrease the logical error rate for a large range of $\theta$, however at certain points the error suppression (e.g. between $0.15\pi$ and $0.225\pi$) was minimal. Therefore we could not confidently pin-down a threshold for systematic rotation. 
\section{Optimizations}
\label{s:optimizations}
A number of optimizations were required for our exact algorithm to handle the code sizes studied in this work. In this section, we will briefly describe the main optimizations used. 

\medskip
\noindent{\em Reducing the inter-column bond dimension --- } Recall that the exact tensor contraction algorithm we use is exponential, which results from the fact that amount of memory required to store the tensor associated to a column is exponential in the number of inter-column bonds. Every check projector that is applied in the algorithm adds bonds to the tensor network. Therefore we can obtain substantial savings if we can reduce the number of check projectors that are applied. In certain calculations, we do not need to apply all of the checks that we might naively expect. For instance, when computing the logical channel $\mathcal{E}$ as detailed in Sec. \ref{s:logical_channel}, we need to apply the recovery map $\mathcal{R}_s$ to the state. The definition of $\mathcal{R}_s$ involves two layers of checks $\Pi_s$, one to the left and one to the right of the input. However, by making use of the cyclic property of the trace, one of the projectors can be absorbed into the other thereby reducing the number of bonds in the network. 

Also, when simulating $z$-rotation, only $x$-checks ever need to be applied. The $z$-checks commute with the noise, and therefore act as the identity on any state in the code space. This is why it was possible to simulate larger codes for $z$-rotation compared with other noise models. 

\begin{figure}
    \centering
    \begin{subfigure}[b]{0.3\textwidth}
        \includegraphics[width=\textwidth]{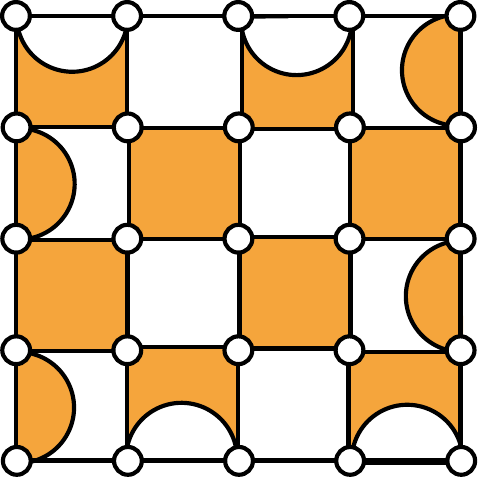}
        \caption{}
    \end{subfigure}\hspace{1cm}%
    \begin{subfigure}[b]{0.3\textwidth}
        \includegraphics[width=\textwidth]{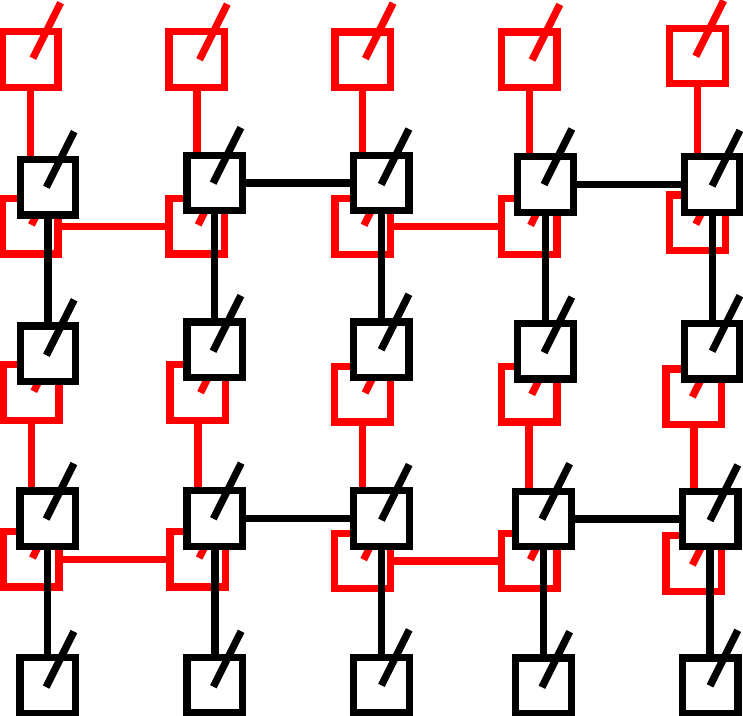}
        \caption{}
    \end{subfigure}
    \caption{a) Distance-5 surface code layout with qubits on the vertices of a rectangular lattice. Orange faces indicate $x$-checks, while white faces indicate $z$-checks. b) A tensor network representing an encoded $\ket{0}_L$ state on this layout. Each box is a $Q^{+}$ tensor, which is applied either to a physical $\ket{0}$ state or to the output leg of another $Q^{+}$ tensor. We have coloured the bottom layer of checks red and the top layer black to make them easier to distinguish. Virtual indices all lie in the plane of the page, and physical indices are perpendicular to it. The bond layout is optimised for the exact contraction scheme we have used.}
     \label{fig:TN}
\end{figure}

Finally, we can reduce the number of inter-column bonds substantially by exploiting the invariance of the check-projector tensors in Eq. \eqref{e:tensor_projector} under permutation of the particles, and always orienting the three bonds of the check such that there is only one inter-column bond per check. This optimised layout is illustrated for the $\ket{0}_L$ state in Fig.~\ref{fig:TN}b). 

\medskip
\noindent{\em Reusing tensor contractions --- } In many cases, when computing the contraction of the square lattice tensor network, the entire network does not need to be contracted. When sampling syndrome measurements, for instance, a check measurement does not affect the tensors in columns to the left and right of the check. The contractions in these unaffected regions can therefore be reused for every check measurement in a column. Similar savings can be obtained when computing different elements of the logical channel process matrix $C_{ij}$, since the application of a logical $Z$ only affects a single column of the tensor network. 

\medskip
\noindent{\em Optimized layout --- } We have used an optimized surface-code layout which was introduced in \cite{bombin_optimal_2007}. This layout achieves a given code distance with fewer physical qubits than the standard layout. It consists of four-qubit $x$ and $z$ checks arranged in a checkerboard pattern, as well as a number of two-qubit checks along the boundaries. We have illustrated this layout in Fig. \ref{fig:TN}a). 
\end{document}